\documentclass[twocolumn]{article}
\usepackage[margin=0.65in]{geometry}
\usepackage[T1]{fontenc}
\usepackage[utf8]{inputenc}
\usepackage{color}
\usepackage{lmodern}
\usepackage{graphicx}
\usepackage[english]{babel}

\usepackage{amsmath}

\usepackage{hyperref}
\hypersetup{
    colorlinks=true,
    linkcolor=black,
    citecolor=black,
    filecolor=black,
    urlcolor=black,
    linkbordercolor = {1 1 1},
}

\usepackage{CJKutf8}

\title{Identifying synergies in private and public transportation}

\author{
	Iva Bojic$^{1,\ast}$, D\'aniel Kondor$^1$, Wei Tu (涂伟)$^2$, Ke Mai (麦可)$^2$, Paolo Santi$^{3,4}$, Carlo Ratti$^3$ \\[2ex]
		\normalsize{$^1$ Singapore-MIT Alliance for Research and Technology, Singapore}\\
		\normalsize{$^2$ Guangdong Key Laboratory of Urban Informatics, Department of Urban Informatics,}\\ \normalsize{School of Architecture and Urban Planning, Shenzhen University, China}\\
		\normalsize{$^3$ Senseable City Laboratory, MIT, Cambridge MA, USA}\\
		\normalsize{$^4$ Istituto di Informatica e Telematica del CNR, Pisa, Italy}\\
		\normalsize{$^\ast$ E-mail: \texttt{ivabojic@mit.edu}}}

\begin{document}

\begin{CJK*}{UTF8}{gbsn}

\maketitle

\end{CJK*}

\section*{Abstract}
In this paper, we explore existing synergies between private and public transportation as provided by taxi and bus services on the level of individual trips. While these modes are typically separated for economic reasons, in a future with shared Autonomous Vehicles (AVs) providing cheap and efficient transportation services, such distinctions will blur. Consequently, optimization based on real-time data will allow exploiting parallels in demand in a dynamic way, such as the proposed approach of the current work. New operational and pricing strategies will then evolve, providing service in a more efficient way and utilizing a dynamic landscape of urban transportation. In the current work, we evaluate existing parallels between individual bus and taxi trips in two Asian cities and show how exploiting these synergies could lead to an increase in transportation service quality.

\section{Introduction}

\begin{figure}
\centering
\includegraphics[width=0.42\textwidth]{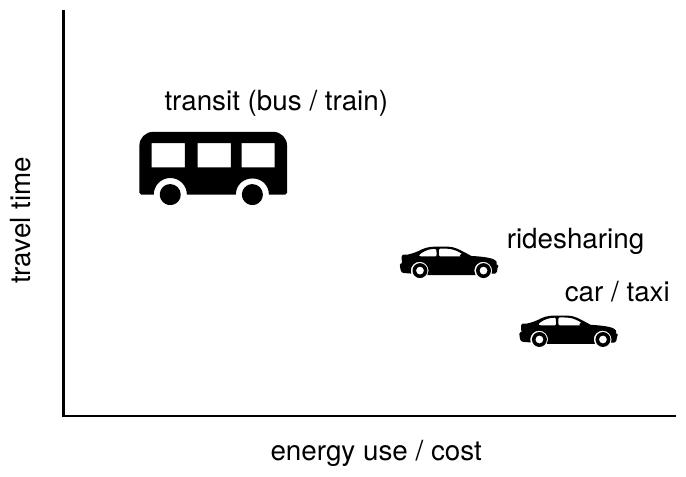}
\caption{Illustration of tradeoffs in travel time, cost and energy use that is typical in today's transportation. Car and bus icons by Rediffusion and Rainbow Designs from the Noun Project, used under CC BY.}
\label{travel_tradeoff}
\end{figure}

In today's transportation industry, there is a large operational gap between private end-to-end services such as taxis\footnote{We note that the taxi market is often highly regulated and in some cases, taxi companies are government-owned. Due to these reasons, some authors consider taxis as part of public transportation. Nevertheless, for the purposes of the current work, we believe the important distinction is along the lines presented here.} and ride-hailing, and fixed-route public transportation such as buses and subways~\cite{Jacques2013, Verbavatz2019}. While the former provide more convenience for passengers and usually a significantly shorter travel time at a higher price, the latter operate on inflexible schedules and routes, with many intermediate stops that slow down the service, but at a significantly lower cost to passengers, operators and society. Looking from a global perspective, we can contrast the total passenger travel time with the total fleet size and vehicle distance traveled. Public transportation modes have a higher travel time, but achieve this with a lower fleet size, total travel distance and energy use. Private transportation modes often have much lower total travel times at the cost of significantly higher fleet size and total energy use, as shown in Fig.~\ref{travel_tradeoff}.

Such tradeoffs are important, as transportation today is a part of every good or service produced \cite{greene1997full}, while commuting takes up an important portion of people's time \cite{ingraham2016astonishing}. This way, the choices for transportation have overarching effects for economic productivity and quality of life \cite{lowrey2011your}. In many developed countries, government policies in the past decades have favored private transportation by ensuring low fuel prices and investing in road infrastructure, despite the increasing evidence of the significant societal cost of private transportation, up to 28 times higher than public transportation~\cite{jakob2006transport}. Due to the large increase of vehicles on roads, cities around the world face significant problems due to congestion; in peak hours, public transportation options can provide a shorter travel time when separated from general traffic~\cite{lindau2010curitiba}. A high usage of private vehicles also contributes to local emissions, while globally, the high total energy use of the transportation sector is also concerning~\cite{shunping2009review}.

The traffic situation is even more problematic in developing countries where underinvestment in both private and public transportation infrastructure resulted in significant congestion and delays for users of both modes. With the lack of public transportation alternatives, the majority of commuters often use private modes, such as in the Klang Valley, Malaysia, where only 17\% of daily trips are made using public transport~\cite{chiu2014mode}, compared to more than 60\% or 90\% of trips made in Singapore and Hong Kong, respectively~\cite{luk2003integrated}.

With the expected availability of Autonomous Vehicles (AVs) in near future~\cite{Fagnant2015, OECDInternationalTransportForum2015, Stead2019}, it will be possible for the above situation to change, with in-between solutions becoming viable~\cite{Christie2016, Alonso-Mora2017}. Eliminating drivers' salary can result in Autonomous Mobility On Demand (AMOD) services becoming dramatically cheaper than services with human drivers today~\cite{Burns2013, Brownell2014}. This raises concerns if public transportation can remain competitive~\cite{Smith2012}. At the same time, these cost savings will allow public transportation operators to explore innovative solutions with new vehicle form factors to provide increased level of service for commuters without increasing costs.

Some of these effects can already be seen with the disruptive changes caused by the emergence of Transportation Network Companies (TNCs), such as Uber, DiDi or Grab~\cite{shaheen2016mobility}. While originally perceived as competitors to taxis~\cite{anderson2014not,gloss2016designing}, it soon became apparent that TNCs serve as competitors to public transportation as well~\cite{feigon2016shared,rayle2016just,hawas2016multi}. Nevertheless, it has been observed that TNCs can also serve a complementary role, especially in areas and time periods that are not served well by current public transportation options~\cite{jin2019uber,grahn2020travelers}.

In addition to TNCs disrupting transportation systems today, Mobility-as-a-Service (MaaS)~\cite{hietanen2014mobility} also brings changes in public transportation by aiming to explicitly integrate public and private transportation modes. The effects of MaaS on public transportation were explored in different countries such as Sweden~\cite{smith2018mobility}, Finland~\cite{heikkila2014mobility} and Germany~\cite{schikofsky2020exploring}. Focus of those research studies was mostly on how to find a ``sweet spot'' that supports innovation, but also secures public benefits. Further research has explored such an integration and collaboration specifically in the context of shared AVs and AMOD services~\cite{Salazar2018, Shen2018}.

In this work, we aim to characterize potential benefits of a more flexible cooperation between private and public transportation services by investigating existing synergies between public and private transportation. Potential applications can depend on the pricing and market strategy of operators and policy decisions of governments to maximize public good. We envision a future where private (i.e.\ non-shared) on-demand transportation remains a ``premium'' option, while a flexible array of shared options provide better convenience, shorter travel times in a more efficient way than current public transportation and ridesharing services can. In this context, our goal is to inform about the shareability potential that can be most easily realized thanks to the reduced cost of AV operations.

More specifically, in this paper, we investigate what are the potential benefits for the passengers who would regularly take bus services if paired with the taxi passengers who take the same route at approximately the same time. With this pairing, the bus passengers can reduce their travel time and the taxi passengers can save money, while the volume of traffic would not increase. There are two main research questions in this paper: (i) what is the percentage of bus passengers who could be matched with the taxi passengers if we assume that both taxi and bus demand is based on today's situation; and (ii) what is the average travel time saved per a bus trip for the passengers who were matched. Answering these questions can give us an understanding about the potential of partial integration of public and private transportation systems. Namely, while not significantly affecting existing taxi passengers or increasing road traffic, our analysis shows a way to improve public transportation experience by reducing a travel time and increasing a travel comfort for its passengers. By showing case studies in two different cities (Singapore and Shenzhen), our analysis also hints at the generalizability of our findings, at least in the context of Asian metropolises.

The rest of the paper is organized as follows. Section~\ref{dataset} gives an overview of the datasets used in our analysis. As we are comparing bus and taxi trips for two different cities, we have four different sources describing people's mobility patterns. Section~\ref{methodology} describes our methodology, formally defining the matching process between trips and explaining the steps of our analysis in detail. Section~\ref{results} shows our main findings answering the two research questions we pose in this paper, i.e.~the percentage of bus trips that can be matched and the average travel time saved for those matched trips. Finally, Section~\ref{discussion} discusses our results and shows directions for future work.

\section{Data}
\label{dataset}

\begin{figure*}
\centering
\includegraphics[width=0.85\textwidth]{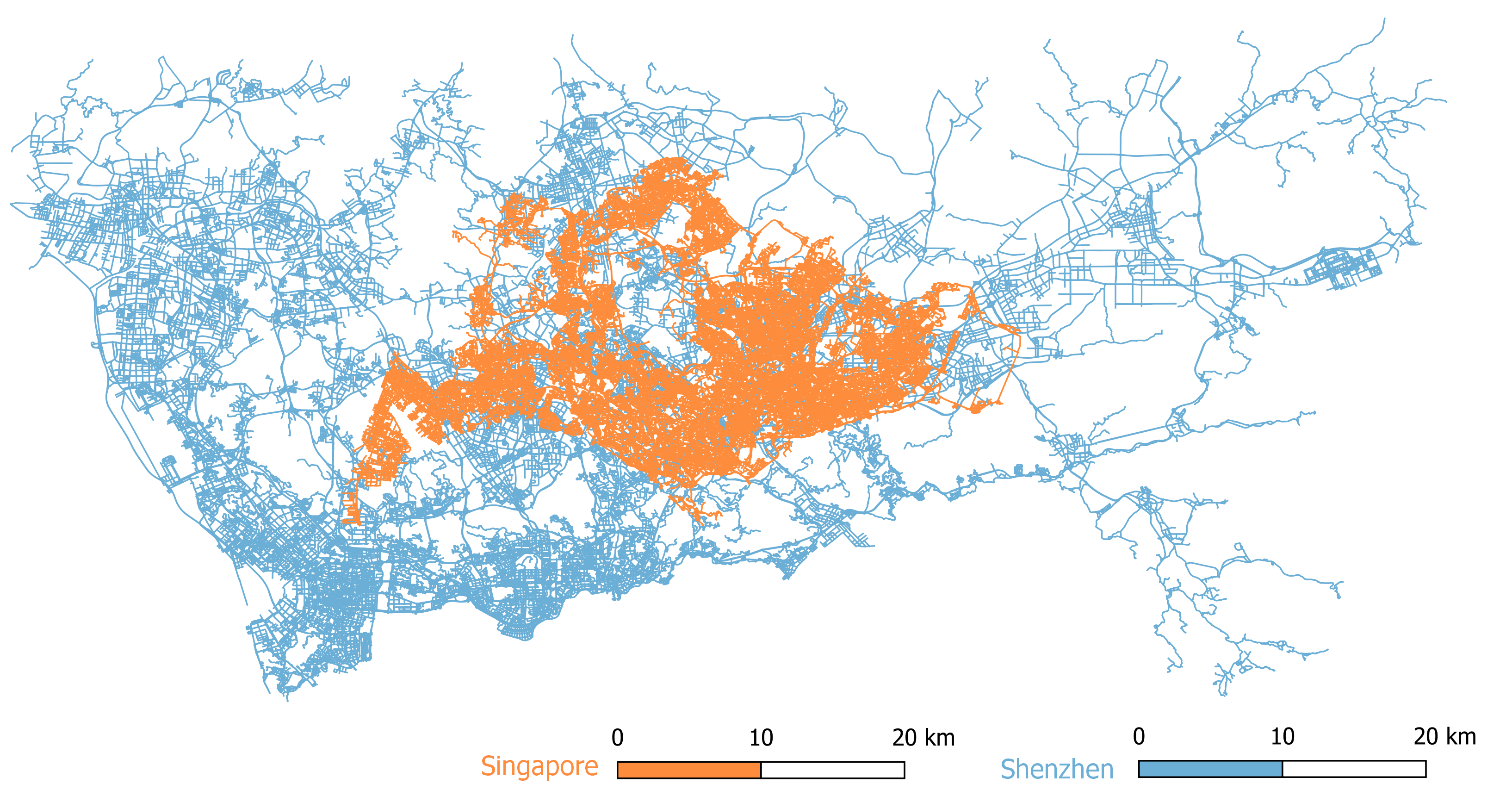}
\caption{The road networks for Singapore and Shenzhen displayed on the same scale.}
\label{Fig4}
\end{figure*}

In this paper, we compare the percentage of bus trips that can be shared with taxi trips in two cities, namely Singapore and Shenzhen. The taxi dataset for Singapore consists of more than 30 million of taxi trips recorded for 77 consecutive days, a total of 10 weeks. Distribution of the number of taxi trips per each day is given in SI Fig.~\ref{FigS1}, where each day in a week is marked with a different color. What we can observe from the figure is that the number of taxi trips is always the largest on Friday and the smallest on Sunday, while between Sunday and Friday the number gradually grows. What is also evident is that there is a little of variance across the observed weeks, with an exception of the second to last week, where Thursday was a public holiday and consequently the number of taxi trips on the preceding Wednesday was higher and on Thursday itself it was significantly lower. From those 10 weeks, we chose the second week as a representative week for further analysis. In Shenzhen, we used one week of taxi trips as well, that has 3 million trips in total, while in Singapore, we have about 2.8 million trips in the chosen week. Neither dataset includes any personal data of passengers.

Bus trips for Shenzhen were collected during the same week as taxi trips and include tap-in data for 15 million trips in total, made by 6.5 million anonymized passengers who are represented by randomly generated IDs used throughout the data collection period. Using the algorithm developed by Tu~et~al.~\cite{tu2018spatial}, we were able to infer destinations for a total of 3.4~million trips that we use in the following analysis (for a more detailed description, see the Supplementary Material). Bus trips for Singapore were generated based on the aggregate counts of bus usage available in the DataMall public interface~\cite{datamall}. The data is based on smart card tap-in and tap-out events, and includes the hourly total number of bus trips made between any two bus stops in the city over the course of a month, separately counted for weekdays and weekends. In total, we have generated a bit more than 20 million bus trips for one week. We generated daily numbers based on this, assuming a Poisson distribution of individual daily counts and assigned trips in the hourly intervals based on a Poisson distribution of bus arrivals as described in more detail in the Supplementary Material. 

In addition to the taxi and bus datasets, we also downloaded the road networks for the two cities from OpenStreetMap~\cite{haklay2008openstreetmap}. For Singapore, we used a bounding box that covers the whole island, and then manually excluded roads that provide connections to Malaysia, and finally, kept the largest connected component of the resulting road network, yielding the network shown in Fig.~\ref{Fig4}. For Shenzhen, we used a bounding box that covers the official boundary of Shenzhen and then removed the connections to Hong Kong and kept the largest connected component, similarly to the case of Singapore. We further processed the raw networks by performing a friend-of-friend clustering, grouping together nodes with a threshold radius of $20\,\mathrm{m}$, reducing the network size to simplify processing and remove uncertainties from small errors in GPS data. After the clustering procedure, the road network for Singapore had more than 50,000 nodes and 120,000 edges, while the one for Shenzhen had almost 40,000 nodes and 100,000 edges. 

When comparing the numbers of records in Singapore and Shenzhen datasets, what we can conclude is that the number of taxi trips for both cities is comparable, i.e.~around 3 million, with Singapore recording a bit less. However, due to limitation of Shenzhen bus dataset not including tap-out data, out of around 15 million records initially recorded in the dataset, we were able to use only a bit more than 25\%. With that being said, what we can see is a large difference in the ratio of bus trips recorded in each city, with Singapore having around six times more bus trips than Shenzhen. However, on the other hand, in Shenzhen dataset we have individual trips with the exact starting time, which is information that is missing from Singapore dataset as there we only have aggregated numbers of people traveling between each pair of origin/destination bus stops within one hour. When comparing the numbers of nodes in each city's road network, we note that the area of Shenzhen is $2.8$ larger than Singapore; this indicates that the road network of Shenzhen is more sparse, which is also evident in Fig.~\ref{Fig4}.


\section{Methodology}
\label{methodology}

The taxi trip datasets for both Singapore and Shenzhen are in the same format of Global Positioning System (GPS) traces. As taxis move through the city, their geolocation (i.e.~latitude and longitude) is recorded at irregular time intervals. In that sense, for each taxi trip in the dataset, there are multiple spatio-temporal points allowing us to reconstruct the route that it took. As the first step, we mapped the taxi trajectories to the road network, using the algorithm of Yang et al.~\cite{yang2018fastmapmatching}. The result of this procedure is an ordered list of network nodes that are present in the most likely trajectory corresponding to the trip. The advantage of this method is that we are not limited by the irregularity in recording GPS points, and we can thus identify all possible matching opportunities. A taxi trip $T_i$ is then represented as an ordered set of tuples $(n_{ij}, t_{ij})$, $j = 1,2\ldots N_i$, where $n_{ij}$ denotes the sequence of road network nodes identified as part of the trajectory, $t_{ij}$ are the estimated timestamps for each node based on the GPS timestamps and finally $N_i$ denotes the total number road network nodes in a trajectory $T_i$.

For each bus trip, we assign a set of road network nodes as \emph{candidate sets} for the beginning and end of the trip, based on their proximity to the coordinates of the bus stop. This way, a bus trip $B_i$ is represented as the following: $B_i = \{ S_i, t_{i,s}, E_i, t_{i,e} \}$, where $S_i$ and $E_i$ are the candidate sets for the start and end of the trip respectively, and $t_{i,s}$ and $t_{i,e}$ are the estimated start and end times of the trip. We control the selection of the candidate sets with the parameter $d$ that we refer to as the \emph{spatial buffer}. For the special value of $d = 0$, the candidate sets only include the closest node to the bus stop. For $d > 0$, the candidate sets include all road network nodes within an Euclidean distance of $d$. In practice, we use $d = 0, 100\,\mathrm{m}$ and $200\,\mathrm{m}$. The significance of $d$ is to allow a match where the pick-up does not exactly take place at the bus stop. Since the actual start of a passenger's trip is typically not the exact location of a bus stop, this buffer is interpreted in the sense that instead of going to the bus stop, a passenger would walk to a pick-up location that is within an acceptable distance of their original location.

We then compile a set of \emph{potential matches} between bus and taxi trips, $\widetilde{M}$, as pairs of trips where a taxi trip includes road network nodes from the start and the end candidate set of a bus trip in the correct order. We also require that the node in the bus start set is visited by the taxi within a short time interval, $t_B$, defined as a \emph{time buffer} within the start of the bus trip and that the end node is visited earlier than the end of the bus trip, allowing time savings for the bus passenger. Formally, we define:
\begin{multline}
	\widetilde{M} = \{ (T_i, B_j, \tau_{ij})\}\quad \forall (i,j),\quad\textrm{where} \quad \exists \, k,l \\ \textrm{such that}
		\left \{ \begin{array}{ll}
			(1) & 1 \leq k < l \leq N_i \\
			(2) & n_{ik} \in S_j \textrm{ and } n_{il} \in E_j \\
			(3) & | t_{ik} - t_{j,s} | < t_B \\
			(4) & t_{j,e} - t_{il} > 0 \\
			(5) & \tau_{ij} \equiv (t_{j,e} - t_{j,s}) - (t_{il} - t_{ik}) > 0
		\end{array} \right .
	\label{eq:matching}
\end{multline}

The conditions listed here guarantee that (1) road network nodes are visited in the correct order; (2) both the start and the end of the bus trip are visited by the taxi trip; (3) the bus passenger can take the taxi within a $t_B$ temporal buffer of the start of their original trip; (4) they arrive earlier than with their original bus trip; and (5) that the actual travel time is shorter, where we define $\tau_{ij}$ as the travel time saving achieved. Note that this matching procedure does not require a bus trip origin/destination (O/D) pair to be exactly the same as a taxi trip O/D pair in order for two trips to be matched, but simply that a subset of the taxi trip matches the bus trip O/D pair. 

The selection of the time buffer $t_B$ assumes that a passenger would go to the bus stop $t_B$ time before boarding the bus and is willing to wait up to $t_B$ after the original departure time if matched with a taxi trip that still arrives earlier at their destination. Of course, in a more detailed model, $t_B$ could be selected on a per-trip basis, if an estimate of the actual waiting time for the bus and individual tolerance for extra waiting time could be established. The travel time saving $\tau_{ij}$ is defined as the \emph{actual} travel time saving, i.e.~how much faster the trip is realized with the taxi than the original bus trip. Notably, we do not include any time savings due to the taxi trip starting earlier. This likely underestimates the \emph{total} time savings achievable to passengers, but since we do not have a reliable estimate of waiting times for bus passengers, we chose to only focus on the part of the trip spent traveling. We acknowledge that minimizing waiting time could be an important additional goal of any combined on-demand mobility service.

Pairs in $\widetilde{M}$ represent all possible sharing opportunities and form a bipartite graph, where the $\tau_{ij}$ time savings are interpreted as edge weights. Potentially, any trip will have multiple match candidates (i.e.~will be present with a degree $>1$). Using this graph, we then calculate a maximal weighted matching~\cite{Edmonds1965,Dezso2011} to arrive at an ideal assignment of trips that maximize time saved for bus passengers while respecting the condition that each trip can be matched only at most once. To be able to provide a tractable solution and also to limit inconvenience to taxi passengers, we do not consider the possibility of a taxi trip being matched to multiple bus trips consecutively; if there are multiple such candidates, we choose the one that contributes to maximizing time savings globally.

\section{Results}
\label{results}

\begin{figure*}[t]
	\centering
	\begin{minipage}{14.2cm}
		\includegraphics{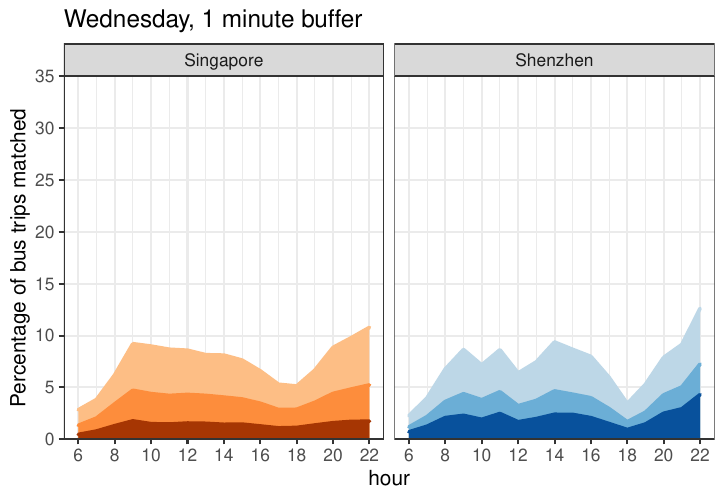}
		\includegraphics{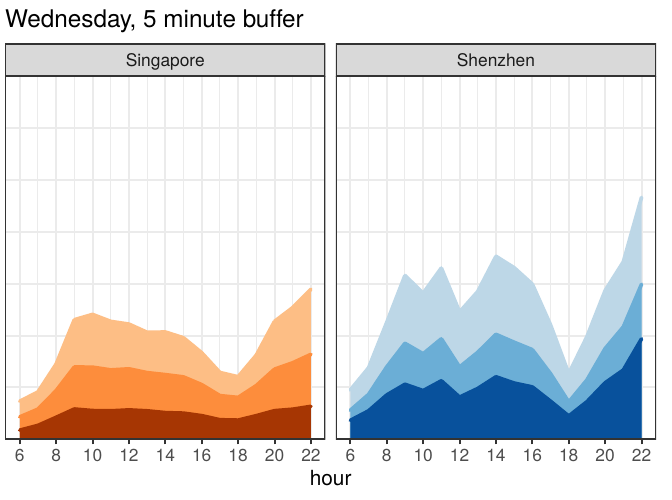}
	\end{minipage}
	\begin{minipage}{1.9cm}
		\includegraphics{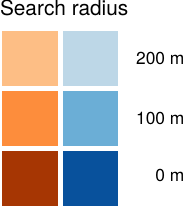}
	\end{minipage} \\
	\begin{minipage}{14.2cm}
		\includegraphics{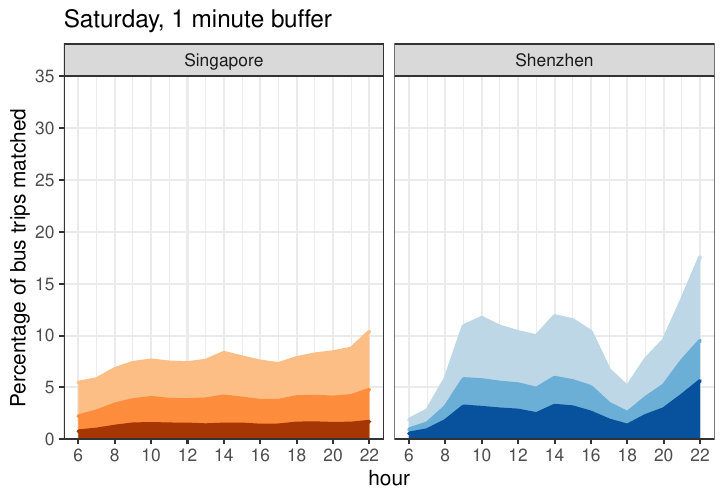}
		\includegraphics{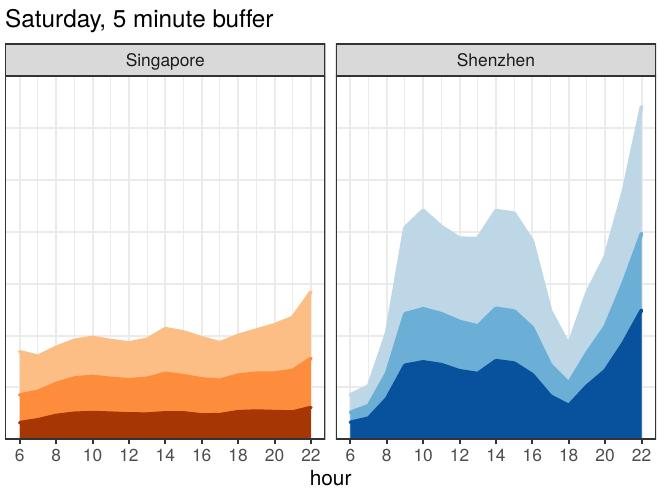}
	\end{minipage}
	\begin{minipage}{1.9cm}
		\includegraphics{legend_vertical}
	\end{minipage}
	\caption{Ratio of bus trips that could be served as a shared trip with a taxi passenger. Results are displayed for Shenzhen and Singapore respectively in the left and right panels. The top row shows results Wednesday and the bottom row shows results for Saturday. Results for the rest of the week are presented in the Supplementary Material, in Figs.~\ref{res:buspctall1} and~\ref{res:buspctall2}. Figures on the left correspond to a time buffer of 1 minute and figures on the right correspond to a time buffer of 5 minutes.}
	\label{res:buspct}
\end{figure*}

\begin{figure*}[t]
	\centering
	\begin{minipage}{14.2cm}
		\includegraphics{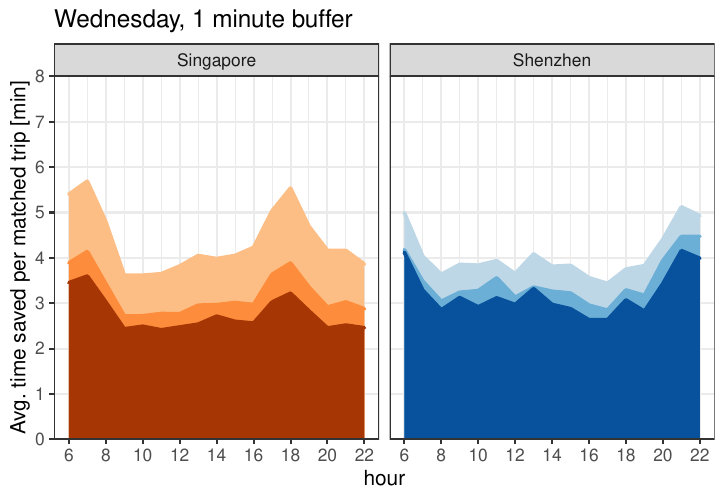}
		\includegraphics{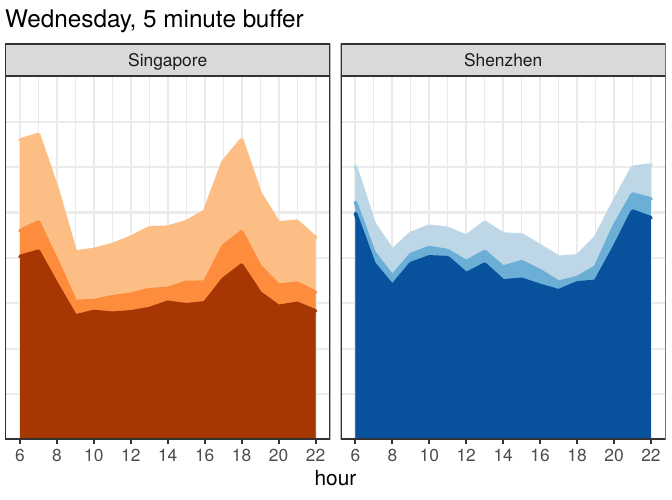}
	\end{minipage}
	\begin{minipage}{1.9cm}
		\includegraphics{legend_vertical}
	\end{minipage} \\
	\begin{minipage}{14.2cm}
		\includegraphics{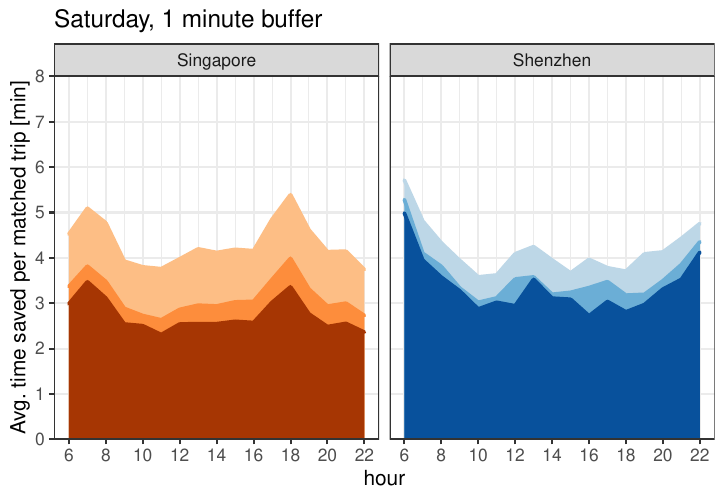}
		\includegraphics{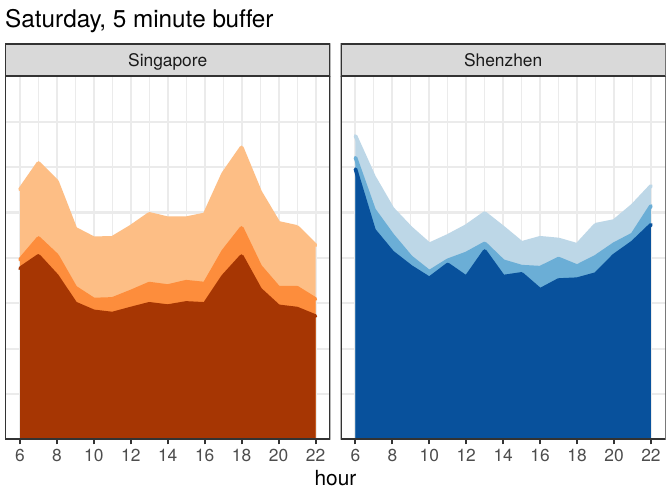}
	\end{minipage}
	\begin{minipage}{1.9cm}
		\includegraphics{legend_vertical}
	\end{minipage}
	\caption{Average time saved by bus passengers who are able to be served in a shared trip. Results are displayed for Shenzhen and Singapore respectively in the left and right panels.  The top row shows results Wednesday and the bottom row shows results for Saturday. Results for the rest of the week are presented in the Supplementary Material, in Figs.~\ref{res:avgtimeall1} and~\ref{res:avgtimeall2}. Figures on the left correspond to a time buffer of 1 minute and figures on the right correspond to a time buffer of 5 minutes.}
	\label{res:avgtime}
\end{figure*}

The results of our analysis are presented with two main figures, each one answering one main research question. Namely, Fig.~\ref{res:buspct} shows the percentage of bus passengers who were able to be matched with taxi passengers, while Fig.~\ref{res:avgtime} shows the average travel time saved per a matched bus trip expressed in minutes. Results for the percentage of matched trips and average time savings are calculated in one hour windows in the time period of significant bus service, between i.e.~6 AM and 11 PM in both cities; correspondingly, $x$-axes are limited between 6 AM and 10 PM. $y$-axes show the percentage of matched trips in Fig.~\ref{res:buspct} and the time savings in minutes in Fig.~\ref{res:avgtime}. Each main figure is divided into eight sub-figures denoting results for Singapore and Shenzhen separately, as well as for different time buffers and for two days (Wednesday and Saturday) that represent typical results for workdays and weekends. Each panel shows results for three different values of the space buffer, i.e.~$d = 0, 100\,\mathrm{m}$ and $200\,\mathrm{m}$. Results for the remaining days of week are displayed in the Supplementary Material as Figs.~\ref{res:buspctall1}~--~\ref{res:avgtimeall2}.

As expected, the percentage of bus trips matched goes up as we increase the space and time buffers. For example, if we set the time buffer to 1 minute, the percentage of matched bus trips on Wednesday for Singapore for the radius of 200 meters is on average a bit less than 10\% in morning hours (i.e.\ between 9 to 12~AM), drops to 5\% around 6~PM and then goes up to a bit more than 10\% in the late night (i.e. around 10~PM). Similar, but slightly lower percentages of matched trips could be also observed for Shenzhen. However, there is a bit of different pattern during the day with two drops around 10 AM and midday and a much sharper drop around 6 PM. The reason for this is the drop of the total number of taxi trips during those periods as shown in SI Fig.~\ref{FigS2} in the Supplementary Material. Namely, whereas the total number of taxi trips in Singapore stays more stable from 10 AM to 1 PM, in Shenzhen there are two drops at 10 AM and midday. Given that the number of bus trips during the same time does not change that much either, chances for the bus passengers to share a ride with the taxi passengers are lower around 10 AM and 12 PM.

For Saturday, the percentage of matched bus tips for Singapore is slightly lower than 10\% and is flatter during the day, with no drop around 6 PM and also rising in late night hours. In Shenzhen, the percentage of bus trips matched on Saturday is slightly larger than 10\%, with two drops around 1 PM and 6 PM and a large spike in late night hours. The reason for that kind of behavior is that there is a little of difference in bus distribution in Shenzhen between Wednesday and Saturday, while the number of taxi trips on Saturday is on average larger than the one on Wednesday. The same patterns are observed for 5 minutes time buffers, but as expected with a higher percentage of bus trips matched. However, the percentage of bus trips matched go much higher for Shenzhen than for Singapore, around 15\% on Wednesday and 20\% on Saturday for 200 meters search radius.

When looking at the average absolute travel time saved for Singapore and Shenzhen (as illustrated in Fig.~\ref{res:avgtime}), what we can see is that the bus passengers in Singapore who get matched with the taxi riders can save a bit more than 4 minutes per their trip for 1 minute time buffer and 200 m space buffer. At the same time, the time savings in Shenzhen (with the same parameters) on average are a bit less than 4 minutes. The time savings with 5 minutes time radius and 200 m space buffer can go over 5 minutes in the case of Singapore and over 4 minutes in the case of Shenzhen.  This also indicates that the increased sharing opportunities contribute to more travel time savings even without including the effect of potentially starting trips earlier that is possible with a larger $t_B$ time buffer. Regarding time patterns, in Singapore we can see larger time savings around 6 AM and 6 PM for both Wednesday and Saturday, while the pattern for Shenzhen is flatter and also does not show a lot of variety between a workday and weekend. This is an interesting finding as when comparing the average bus trip lengths of Singapore and Shenzhen (as shown in SI Fig.~\ref{FigS4}), we can see that on average, bus trips in Singapore are shorter (i.e. mean for Singapore is around 10 minutes and for Shenzhen 14 minutes). This is understandable given that the area of Shenzhen is significantly larger. Consequently, when putting time savings into a perspective, this means that on average, absolute travel time saved in Singapore is up to 50\% of mean bus duration and around 30\% for Shenzhen.

\section{Discussion}
\label{discussion}

Until very recently, private and public transportation have been two systems that were very much separated~\cite{Jacques2013, jakob2006transport}. However, with the emergence of Transportation Network Companies (TNCs) and Mobility-as-a-Service (MaaS) concepts, those lines are becoming more unclear~\cite{shaheen2016mobility, feigon2016shared, jin2019uber, smith2018mobility}. This will be even more evident once Shared Autonomous Vehicles (SAVs) hit the roads bringing further disruptive changes to urban mobility~\cite{Fagnant2015, OECDInternationalTransportForum2015, Stead2019, Smith2012}. If drivers' costs are removed from the equation, vehicles of different, more flexible capacities would be able to operate in public transportation services as well. In practice, this will allow transportation services to be more fluid and will contribute to vehicles being utilized more, e.g.~by introducing services based on shared rides in medium-sized vehicles (6-10 passengers)~\cite{Alonso-Mora2017}, which would provide travel time benefits over public buses and reduced fleet size compared to taxi and ride-hailing services.

In this paper, we thus analyzed the first step of integrating public and private transportation using today's travel demand. Using mobility patterns recorded by taxi companies and bus operators in Singapore and Shenzhen, we investigated how passengers on public transportation can reduce their travel time if paired with already existing taxi riders. In our proposed framework, we can identify three stakeholders - passengers taking public transportation, taxi riders, and the local government being responsible for public roads. The matching concept explored in our work presents a Pareto improvement from the current situation, i.e.~some stakeholders’ experience benefits, while none of them experience losses: the travel time is reduced for some public transportation passengers, the costs are reduced for some taxi passengers, while the total number of cars and traffic on the road does not increase.

The results of our analysis showed that between 10 to 20 percent of bus riders could be potentially matched with taxi riders, which would contribute to on average between 4 to 6 minutes of savings of their travel time. These results are consistent across two cities in Asia, although there are individual differences in the temporal pattern of matching ratios. The main source of difference is explained by how the total volume of taxi and bus trips changes during the day. Namely, first we see a clear difference between the distribution of total bus trips in Singapore between a weekday and a weekend, whereas this difference is less obvious in case of Shenzhen. This possibly means that in Shenzhen, more people also work on Saturdays. Second, the total amount of taxi trips on Saturday in Shenzhen is larger than on Wednesday, which is not true for Singapore.

In conclusion, our analysis shows that there is a practical potential for partial integration of public and private transportation even under the current conditions. This is an important first step when envisioning a future where AV technology allows a variety of novel transportation service types. Future work might look into extensions where bus passengers are not only matched with taxis in an opportunistic manner, but with alternative service providers that can operate with medium sized (i.e.~6-10 passenger) vehicles with the specific goal of providing a faster, more convenient and demand-responsive transportation alternative to buses. We note that previous work in this area was limited to using taxi trips as an estimate of demand~\cite{Alonso-Mora2017}; our results show the importance of including the \emph{complete} picture of urban transportation demand, i.e.~both public and private transportation users. Furthermore, while our work shows a potential for matching trips, any such service will face challenges in implementing user interaction solutions that can be conveniently used without excluding groups of users, e.g.~those who do not use a smartphone. This means that investigating new user interaction concepts for the fluid transportation services of tomorrow will become increasingly important as more and more optimization opportunities in on-demand transportation are realized.

\section*{Acknowledgments}

This research is supported by the Singapore Ministry of National Development and the National Research Foundation, Prime Minister’s Office, under the Singapore-MIT Alliance for Research and Technology (SMART) programme. We also thank RATP, Dover Corporation, Allianz, Teck Resources, Lab Campus, Anas S.p.A., Ford, ENEL Foundation, the Amsterdam Institute for Advanced Metropolitan Solutions, the cities of Laval, Curitiba, Stockholm and Amsterdam, and all of the members of the MIT Senseable City Laboratory Consortium for supporting this research. Wei Tu acknowlegdes the funding of Nature Science Foundation of China (No.4207010598).

\section*{Declarations of interests}

The authors declare no competing interest.

\onecolumn
\newpage
\normalsize

\renewcommand{\thefigure}{S\arabic{figure}}
\renewcommand{\thetable}{S\arabic{table}}

\setcounter{figure}{0}
\setcounter{table}{0}

\begin{center}
{\huge \bf Supplementary Material \\[2.5ex]}
\end{center}

\paragraph{Supplementary Figures}

\begin{figure*}[b]
\centering
\includegraphics{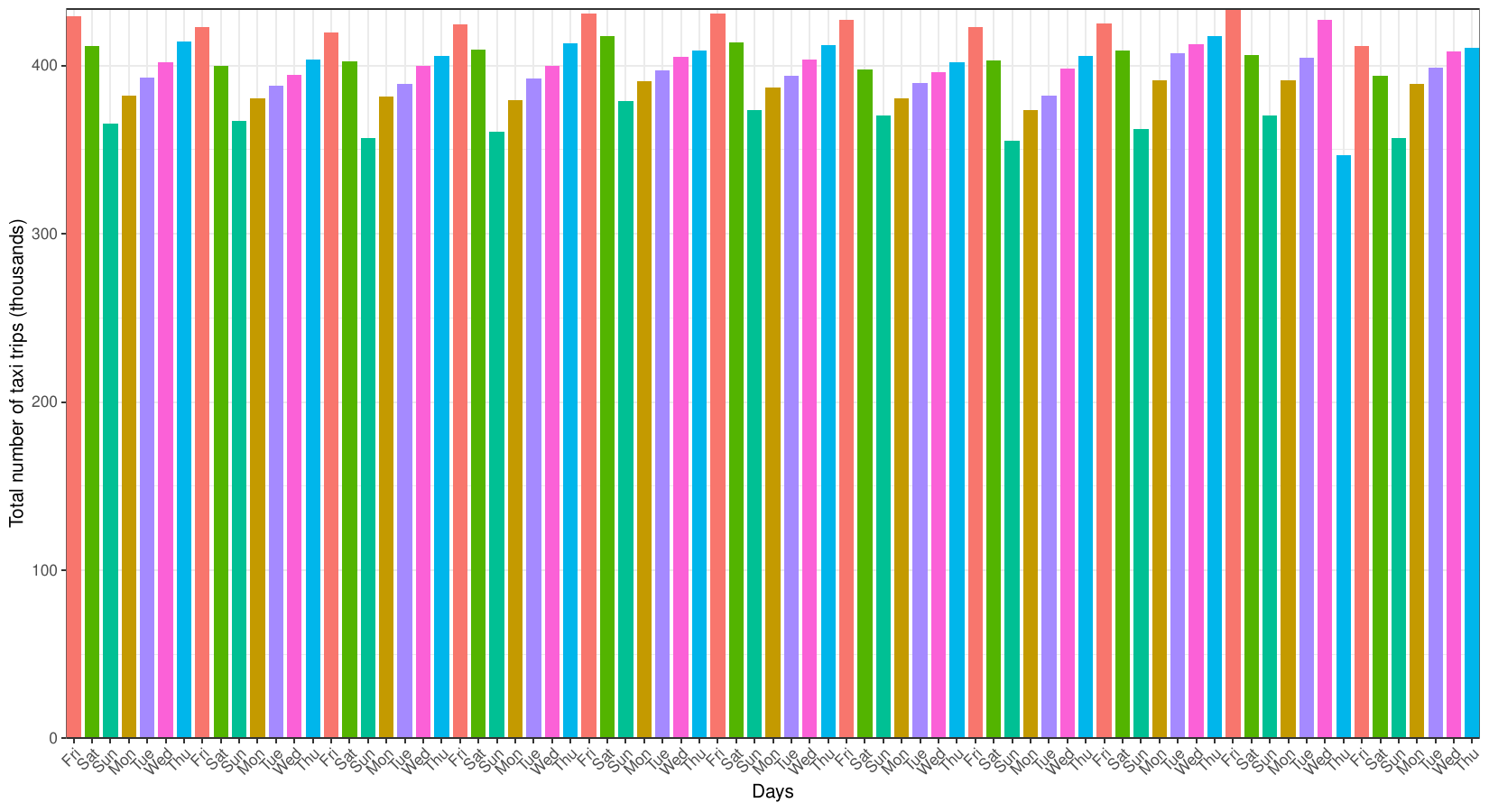}
\caption{The number of taxi trips in Singapore per day.}
\label{FigS1}
\end{figure*}

Figure \ref{FigS1} shows the number of taxi trips in Singapore for each of 77 consecutive days recorded in our dataset.
Figure~\ref{FigS2} shows distribution of the number of taxi and bus trips per hour on Wednesday in Singapore and Shenzhen, while Figure~\ref{FigS3} presents the same distributions, but for Saturday.
Figure~\ref{FigS4} shows distribution of bus trip duration per hour on Wednesday and Saturday in Singapore and Shenzhen.
Figures~\ref{res:buspctall1} and~\ref{res:buspctall2} show the percentage of shareable trips for the rest of the weekdays that were not included in Fig.~\ref{res:buspct} in the main text (i.e.\ Monday, Tuesday, Thursday, Friday and Sunday). Figures~\ref{res:avgtimeall1} and~\ref{res:avgtimeall2} show the average time savings for bus passengers on these days.

\paragraph{Bus trip processing for Shenzhen}

For Shenzhen, we fused bus GPS trajectories and Smart Card Data (SCD) to infer bus trips destinations. We inferred the alighting location and time by considering the spatial-temporal regularity of the SCD user. Firstly, the boarding location of a smart card user was inferred. The continuous bus GPS trajectory was recovered from GPS records by map matching considering the road network and the delay at the road crossing \cite{tu2018spatial}. Each bus SCD record was linked to the corresponding trajectory based on the bus identification number. The recorded time was then used to interpolate the boarding location from the GPS trajectory. Following the direction of the bus route, we adjusted the boarding location to the nearest bus stop in the bus route. Then, the alighting location of the SCD record was inferred. The SCD user with a pair with the highly frequent boarding locations were filtered \cite{zhong2015measuring}. Considering the regularity of commuters, the following highly frequently boarding location was recognized as the alighting location of the preceding SCD record. The alighting time was interpolated according to the corresponding continuous GPS trajectory. 

\paragraph{Bus trip generation for Singapore}

\begin{figure*}
\centering
\includegraphics{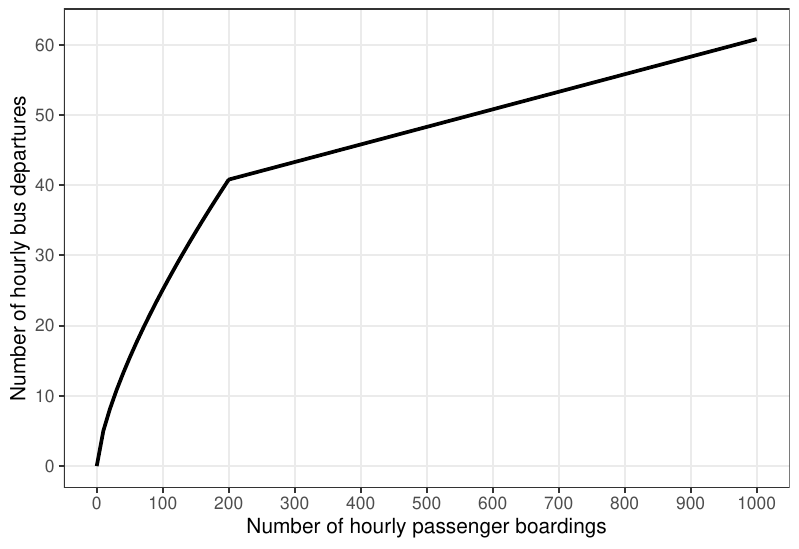}
\caption{Empirically motivated relationship between the hourly number of bus passengers in a stop and the expected number of bus departures, i.e.~the relationship described in Eq.~\ref{eq:nbus}.}
\label{fig:nbus}
\end{figure*}

\begin{figure*}
\centering
\includegraphics{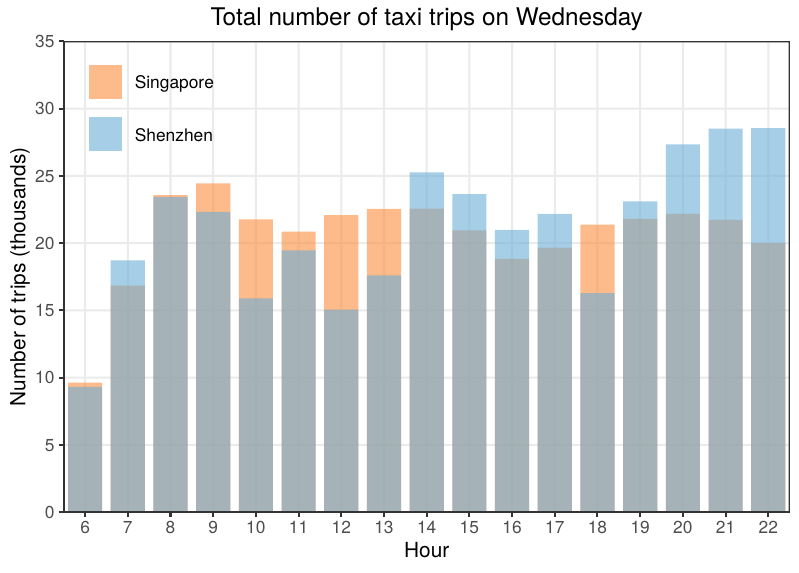}
\includegraphics{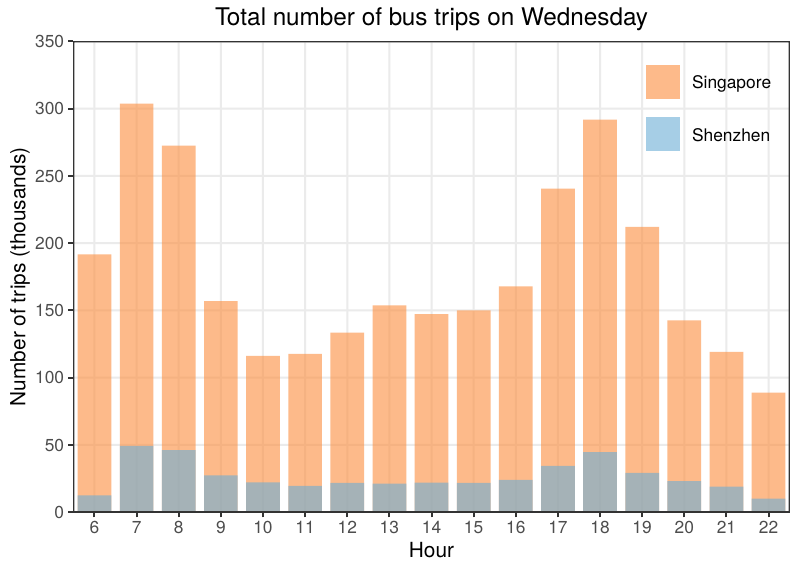}
\caption{Comparing the total number of taxi and bus trips on Wednesday in two cities.}
\label{FigS2}
\end{figure*}

\begin{figure*}
\centering
\includegraphics{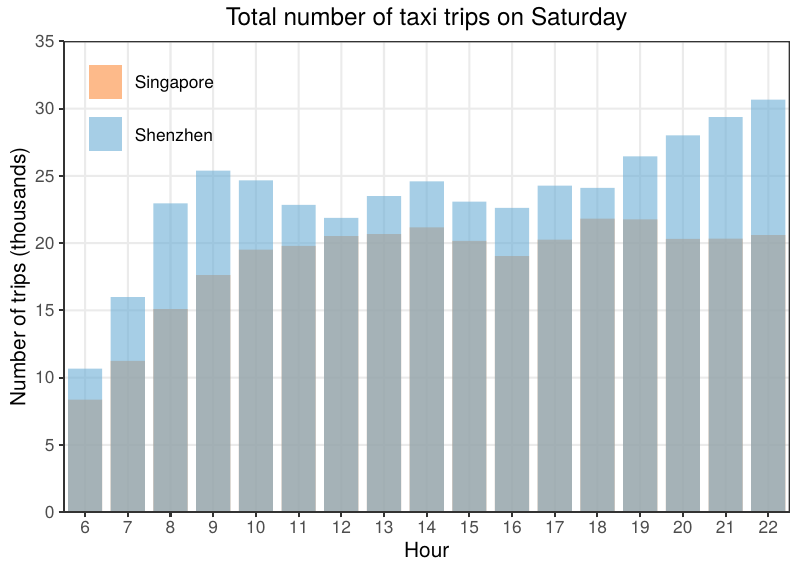}
\includegraphics{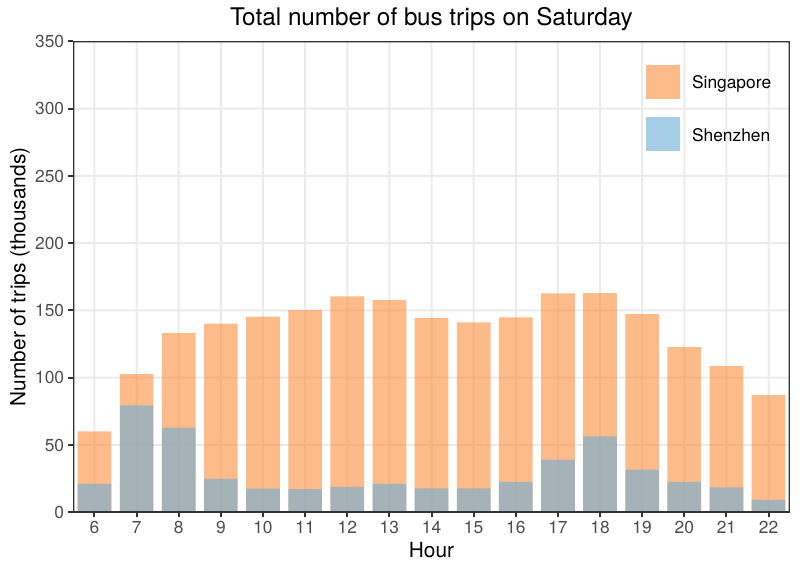}
\caption{Comparing the total number of taxi and bus trips on Saturday in two cities.}
\label{FigS3}
\end{figure*}

\begin{figure*}
\centering
\includegraphics{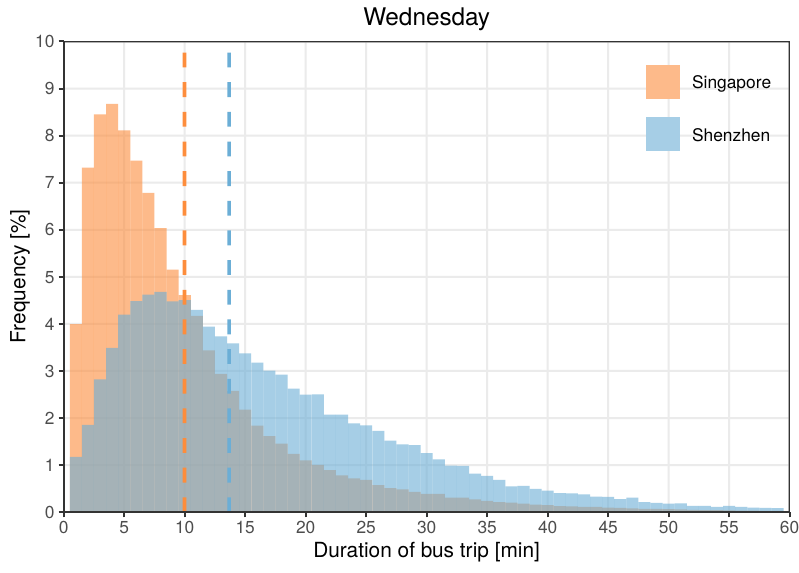}
\includegraphics{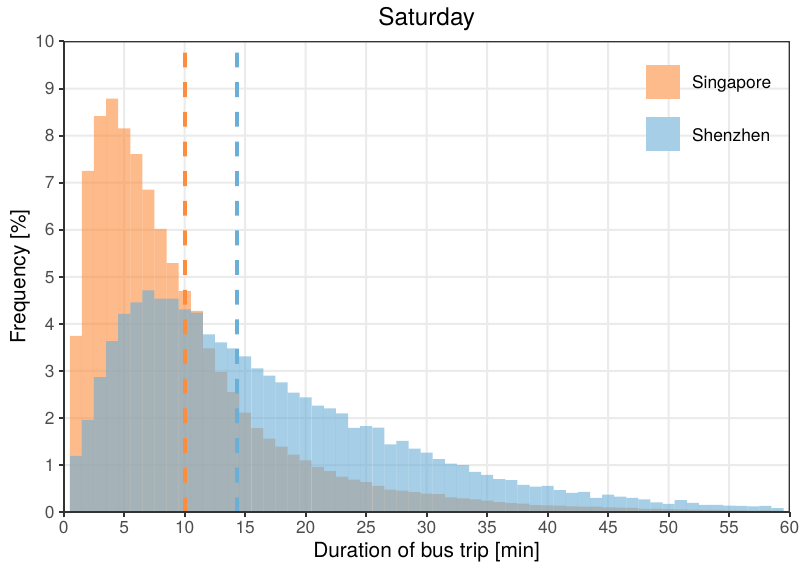}
\caption{Comparing duration of bus trips on Wednesday and Saturday in two cities. Dashed lines display the mean of the distributions, which are $9.98\,\mathrm{min}$ and $10.03\,\mathrm{min}$ in the case of Singapore for Wednesday and Saturday respectively, and $13.67\,\mathrm{min}$ and $14.32\,\mathrm{min}$ in the case of Shenzhen.}
\label{FigS4}
\end{figure*}

\begin{figure*}
\centering
	\begin{minipage}{14.2cm}
		\includegraphics{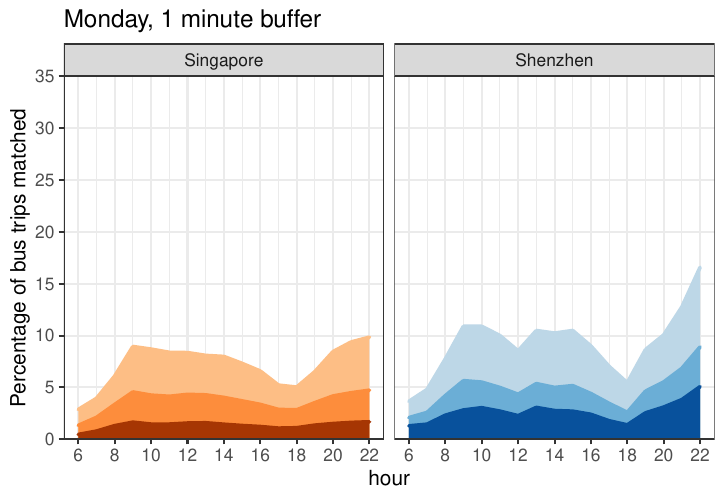}
		\includegraphics{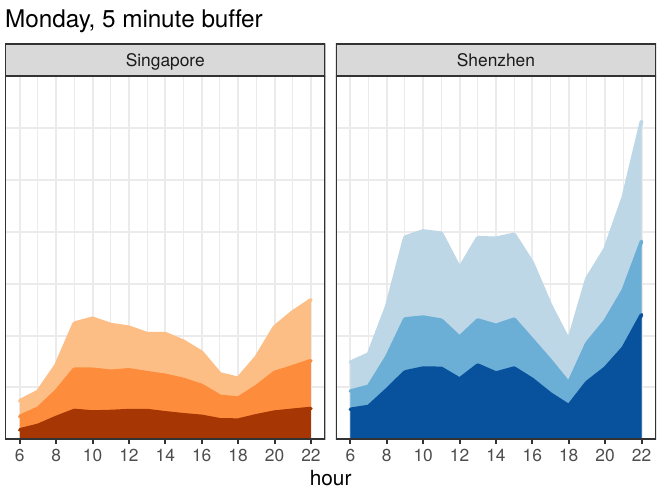}
	\end{minipage}
	\begin{minipage}{1.9cm}
		\includegraphics{legend_vertical}
	\end{minipage} \\
	\begin{minipage}{14.2cm}
		\includegraphics{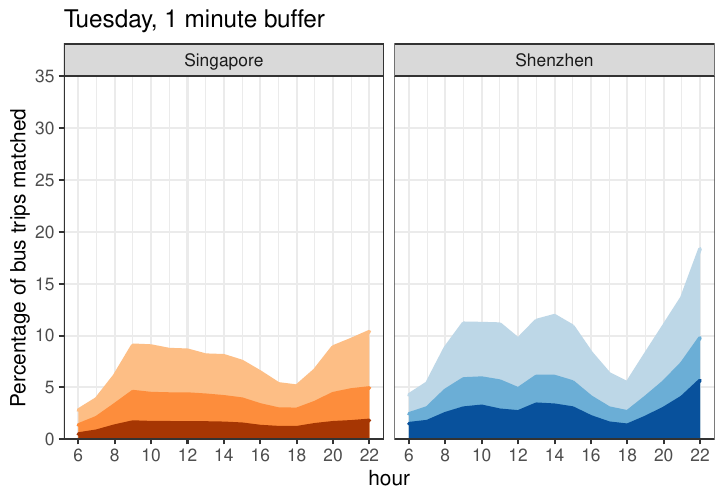}
		\includegraphics{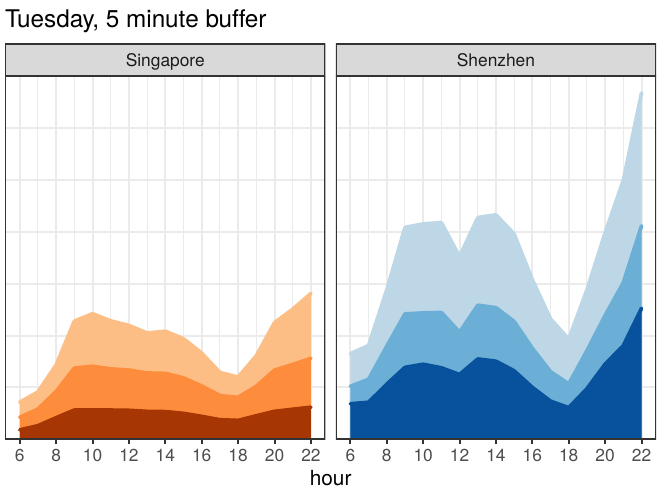}
	\end{minipage}
	\begin{minipage}{1.9cm}
		\includegraphics{legend_vertical}
	\end{minipage} \\
	\begin{minipage}{14.2cm}
		\includegraphics{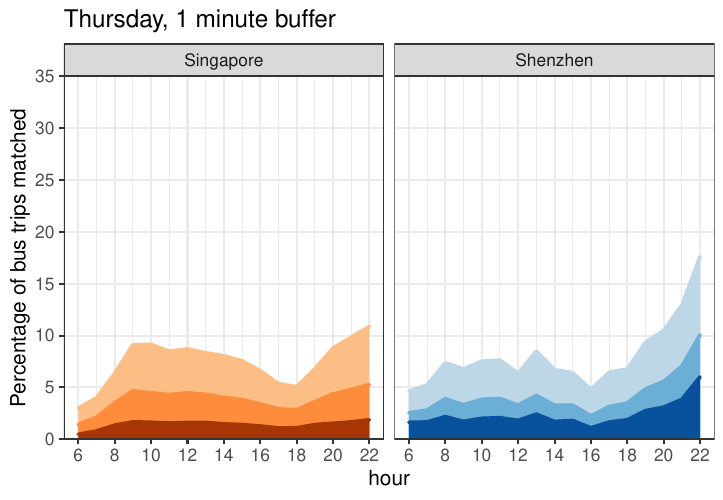}
		\includegraphics{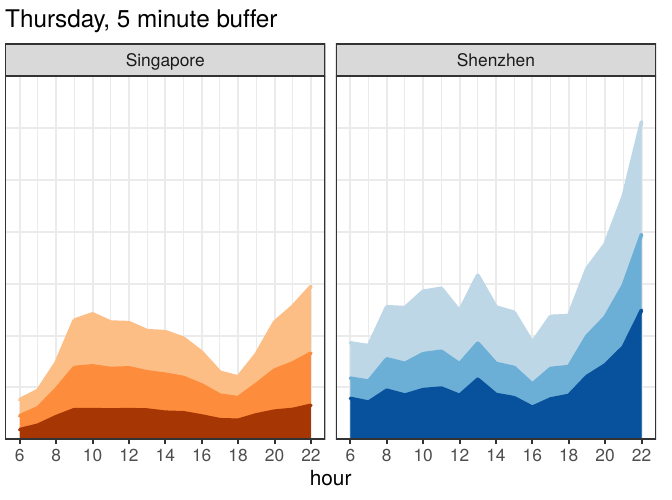}
	\end{minipage}
	\begin{minipage}{1.9cm}
		\includegraphics{legend_vertical}
	\end{minipage} \\
	\begin{minipage}{14.2cm}
		\includegraphics{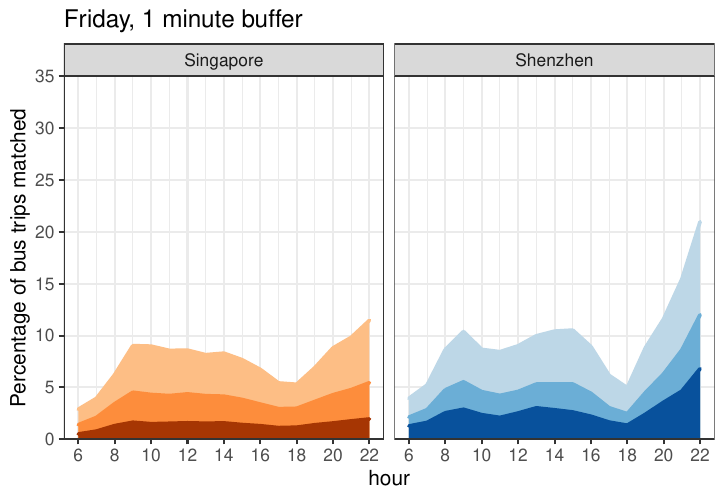}
		\includegraphics{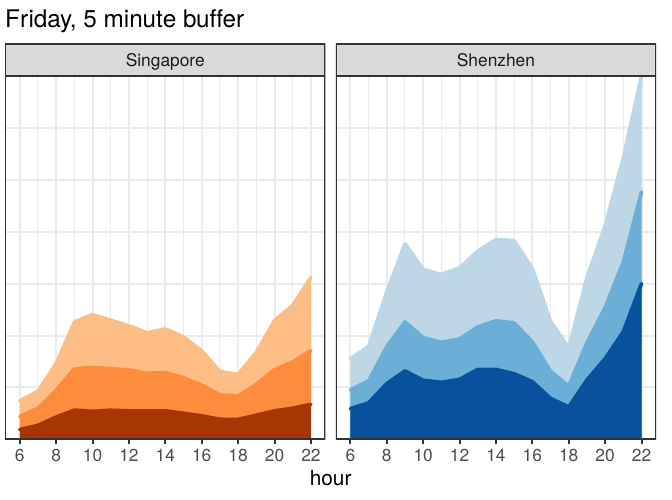}
	\end{minipage}
	\begin{minipage}{1.9cm}
		\includegraphics{legend_vertical}
	\end{minipage}
	\caption{Percentage of bus trips shareable for weekdays.}
	\label{res:buspctall1}
\end{figure*}

\begin{figure*}
\centering
	\begin{minipage}{14.2cm}
		\includegraphics{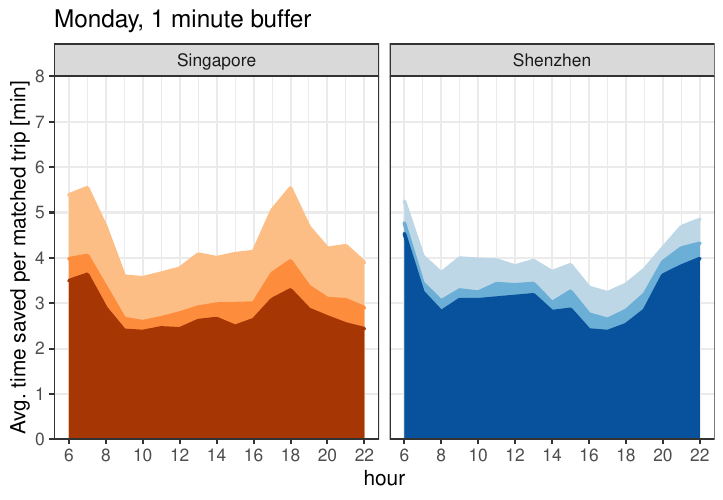}
		\includegraphics{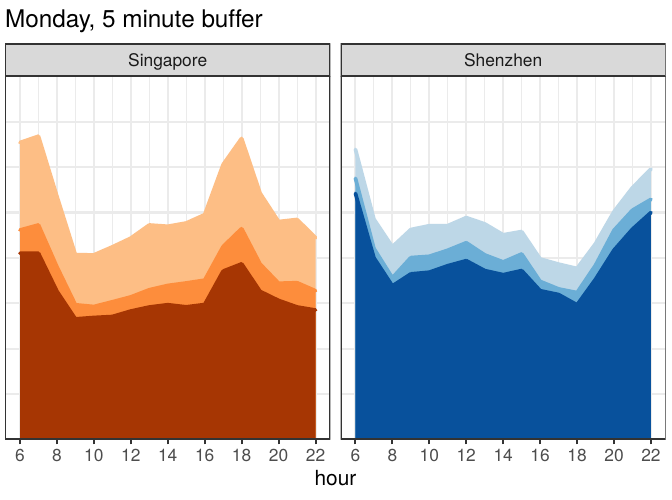}
	\end{minipage}
	\begin{minipage}{1.9cm}
		\includegraphics{legend_vertical}
	\end{minipage} \\
	\begin{minipage}{14.2cm}
		\includegraphics{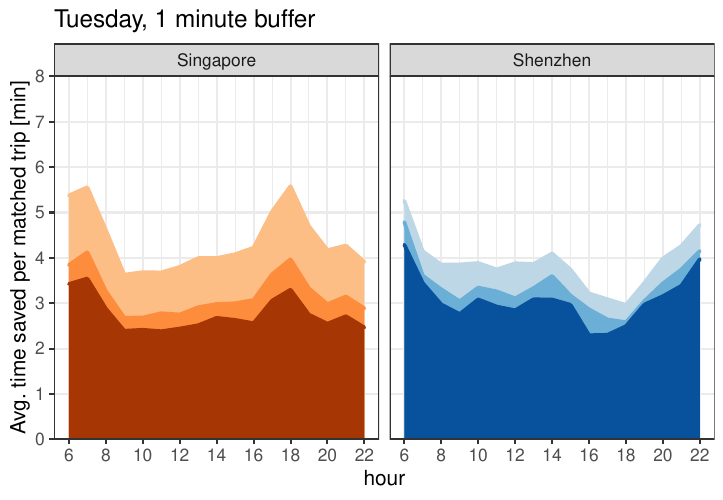}
		\includegraphics{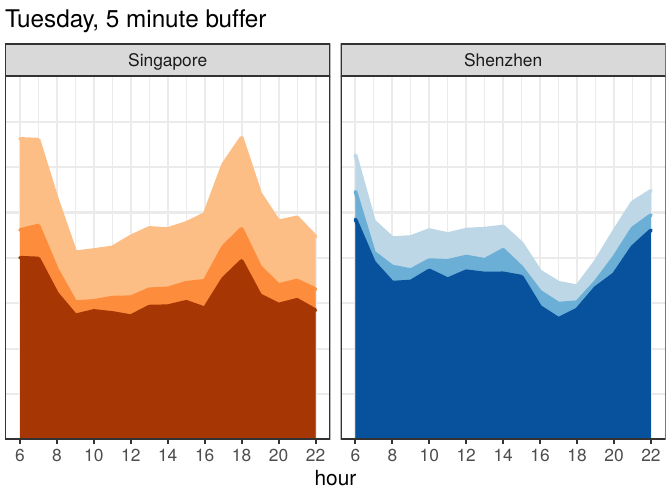}
	\end{minipage}
	\begin{minipage}{1.9cm}
		\includegraphics{legend_vertical}
	\end{minipage} \\
	\begin{minipage}{14.2cm}
		\includegraphics{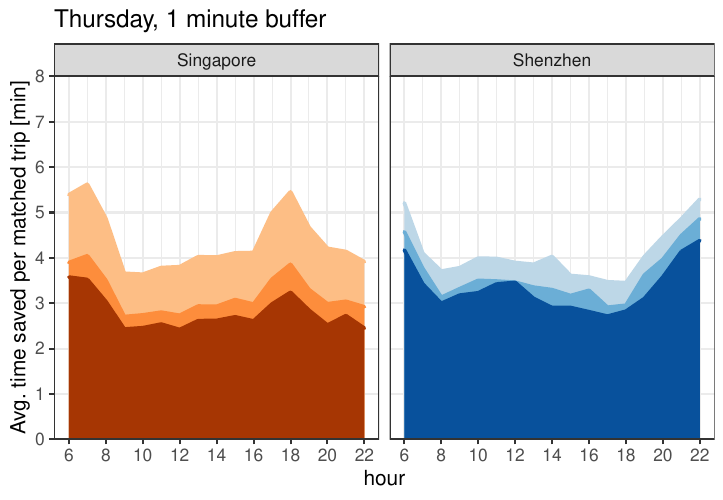}
		\includegraphics{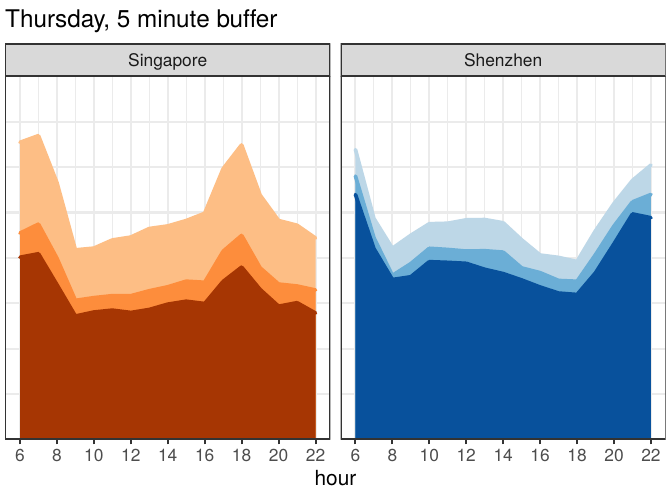}
	\end{minipage}
	\begin{minipage}{1.9cm}
		\includegraphics{legend_vertical}
	\end{minipage} \\
	\begin{minipage}{14.2cm}
		\includegraphics{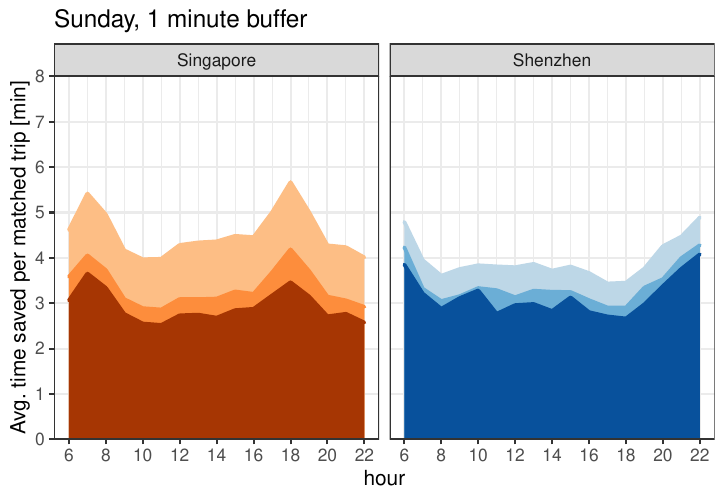}
		\includegraphics{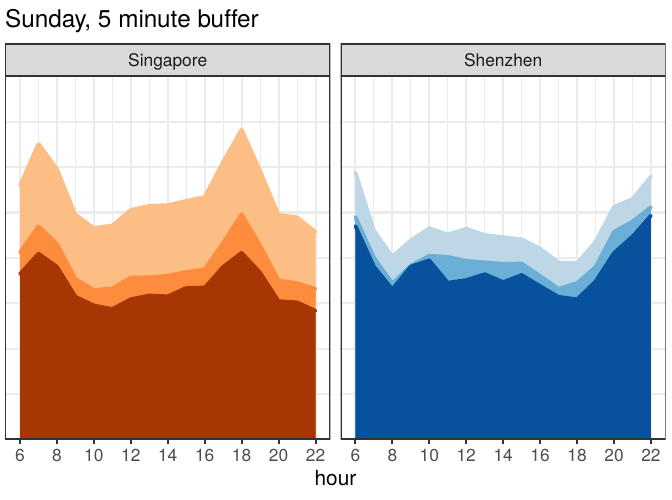}
	\end{minipage}
	\begin{minipage}{1.9cm}
		\includegraphics{legend_vertical}
	\end{minipage}
	\caption{Average time saved for shareable trips on weekdays.}
	\label{res:avgtimeall1}
\end{figure*}

\begin{figure*}
\centering
	\begin{minipage}{14.2cm}
		\includegraphics{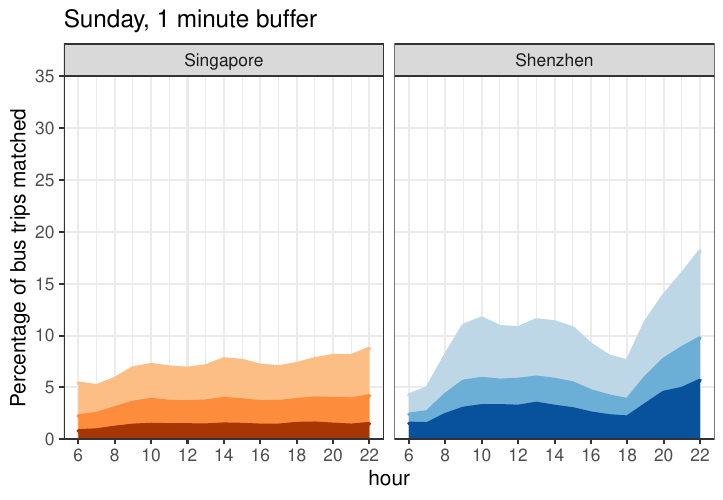}
		\includegraphics{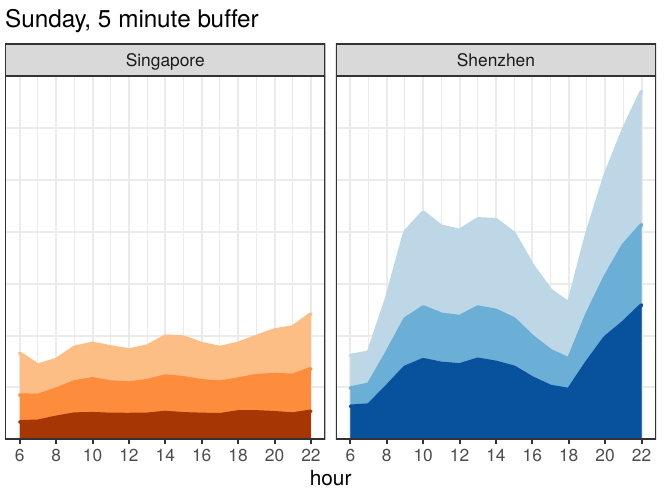}
	\end{minipage}
	\begin{minipage}{1.9cm}
		\includegraphics{legend_vertical}
	\end{minipage}
	\caption{Percentage of bus trips shareable for Sunday.}
	\label{res:buspctall2}
\end{figure*}

\begin{figure*}
\centering
	\begin{minipage}{14.2cm}
		\includegraphics{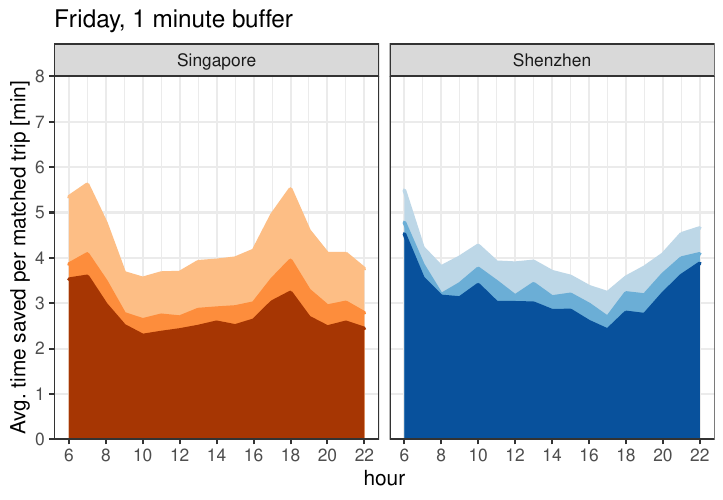}
		\includegraphics{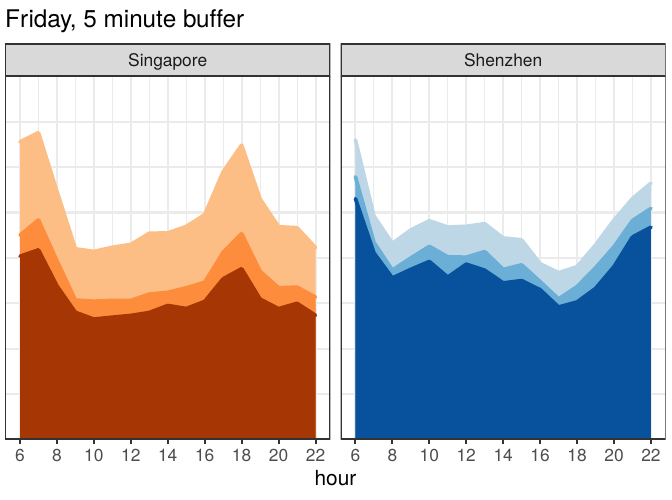}
	\end{minipage}
	\begin{minipage}{1.9cm}
		\includegraphics{legend_vertical}
	\end{minipage}
	\caption{Average time saved for shareable trips on Sunday.}
	\label{res:avgtimeall2}
\end{figure*}

The bus dataset includes hourly counts of trips between any two bus stops in Singapore over the course of one month, aggregated separately for weekdays and weekends. As a first step, we calculated the average hourly counts for each day, by dividing the total numbers with the number of weekdays and weekend days (including public holidays), i.e.~22 and 9 respectively. Next, we generated realizations of the actual number of bus trips between each bus stop pair by sampling from a Poisson distribution with the mean given by the average counts. We repeated this process for each bus stop pair and for each hour of the day when buses are operating. We then distributed trip start times in the one-hour intervals based on a process where a given number of individually distributed bus departures were assumed for each bus stop in each hour. To achieve this, we counted the total hourly passenger counts for each bus stop, denoting by $N_{ij}$ the number of passengers boarding a bus at the $i$th stop of the $j$th hour. We then assumed that the expected number of bus departures in a stop was related to the number of passenger boarding in the following way:
\begin{equation}
	B_{ij} = \left \{ \begin{array}{ll}
		N_{ij}^\alpha \quad \textrm{if} \quad N_{ij} < N^* \\
		B^* + (N_{ij} - N^*) / b_0 \quad \textrm{if} \quad N_{ij} \geq N^*
	\end{array} \right .
	\label{eq:nbus}
\end{equation}
Note that we defined the relation in Eq.~\ref{eq:nbus} based on our previous work with the bus travel data. In Eq.~\ref{eq:nbus}, we used $B^* \equiv (N^*)^\alpha$ and the numerical parameters were $\alpha = 0.7$, $N^* = 200$ and $b_0 = 40$. We displayed the relationship between $B_{ij}$ and $N_{ij}$ in Fig.~\ref{fig:nbus}. Using this relationship, we assigned $B_{ij}$ as the expected number of bus departures for every bus stop and hour and then generated an actual number of departures from a Poisson distribution with $B_{ij}$ mean and excluding the case when this random choice would give a zero value. As the last step, we distributed departure times within the one-hour interval among the bus departures using an exponential distribution and normalizing the total elapsed time and assigned each passenger randomly among the buses, using the departure time of the selected bus as the trip start time.

\end{document}